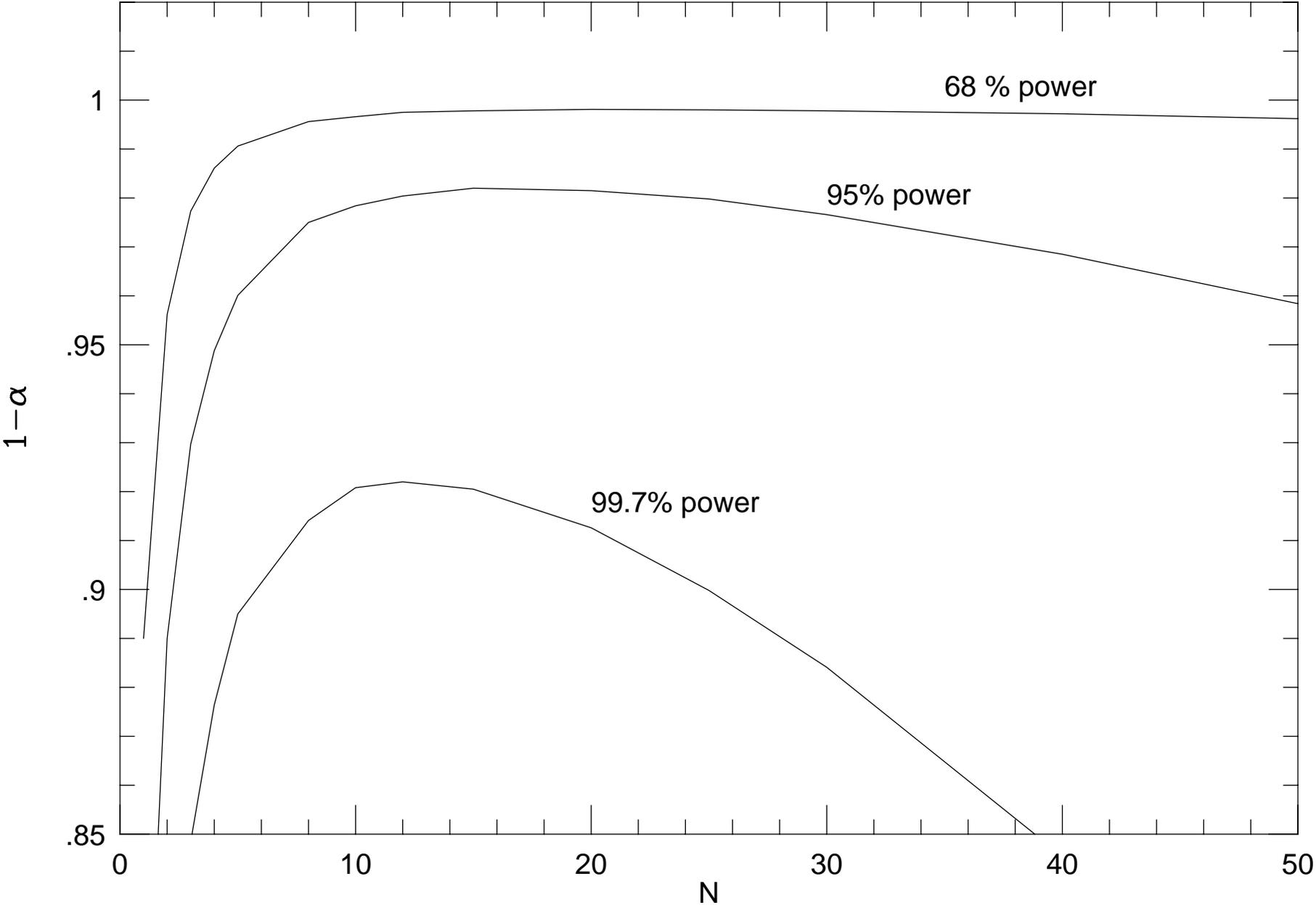

confidence level as a function of number of fields R=7.

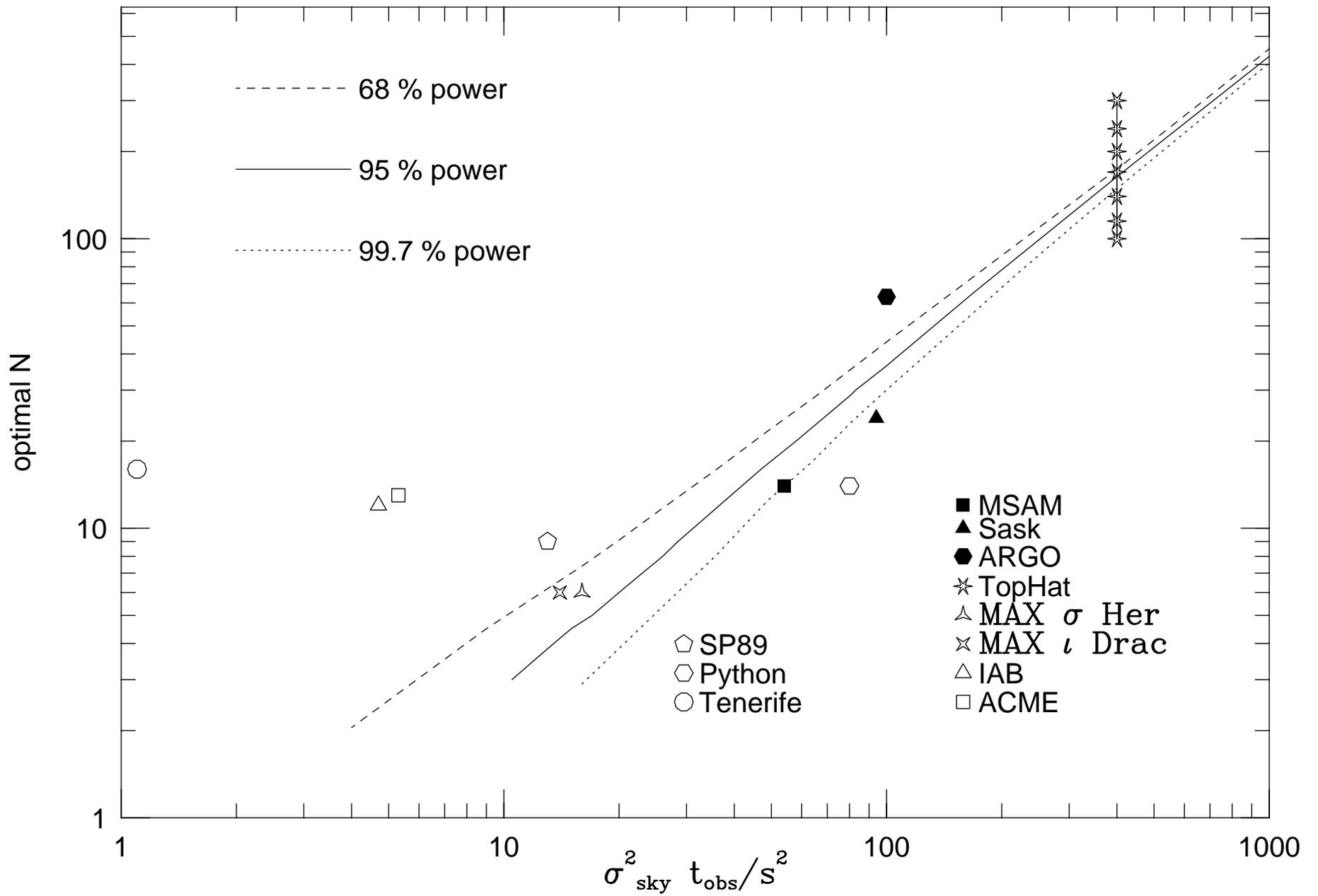

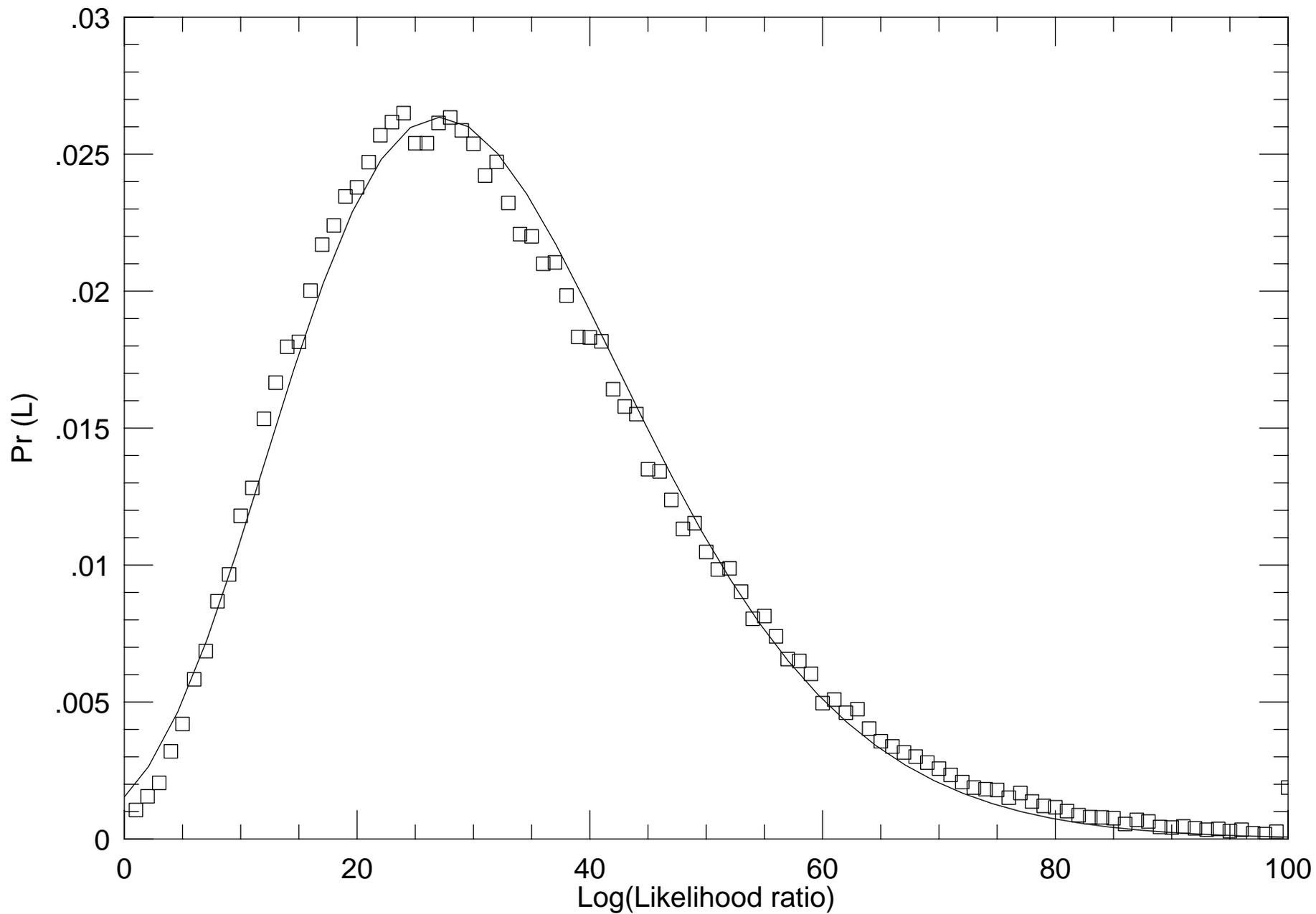

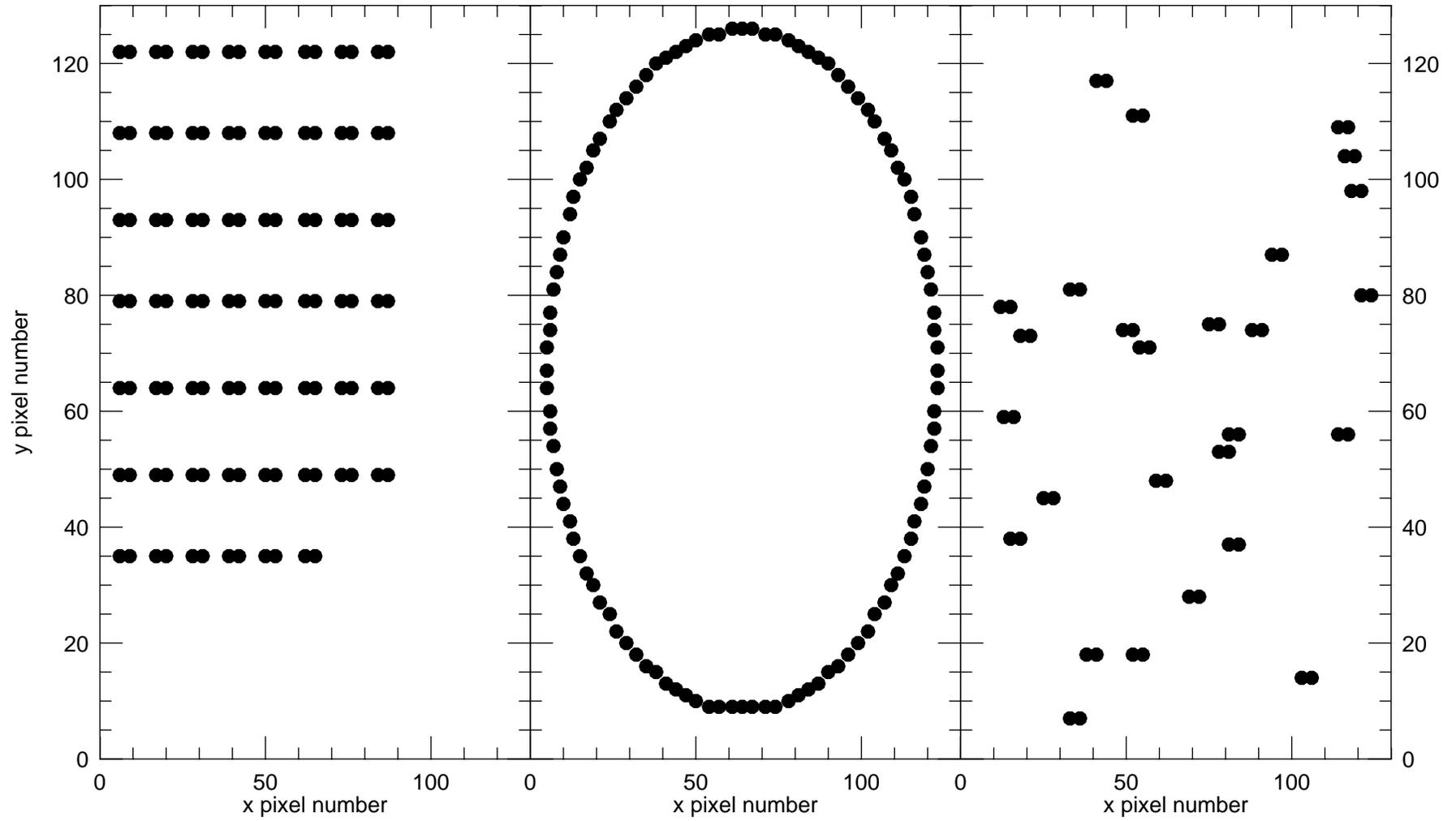

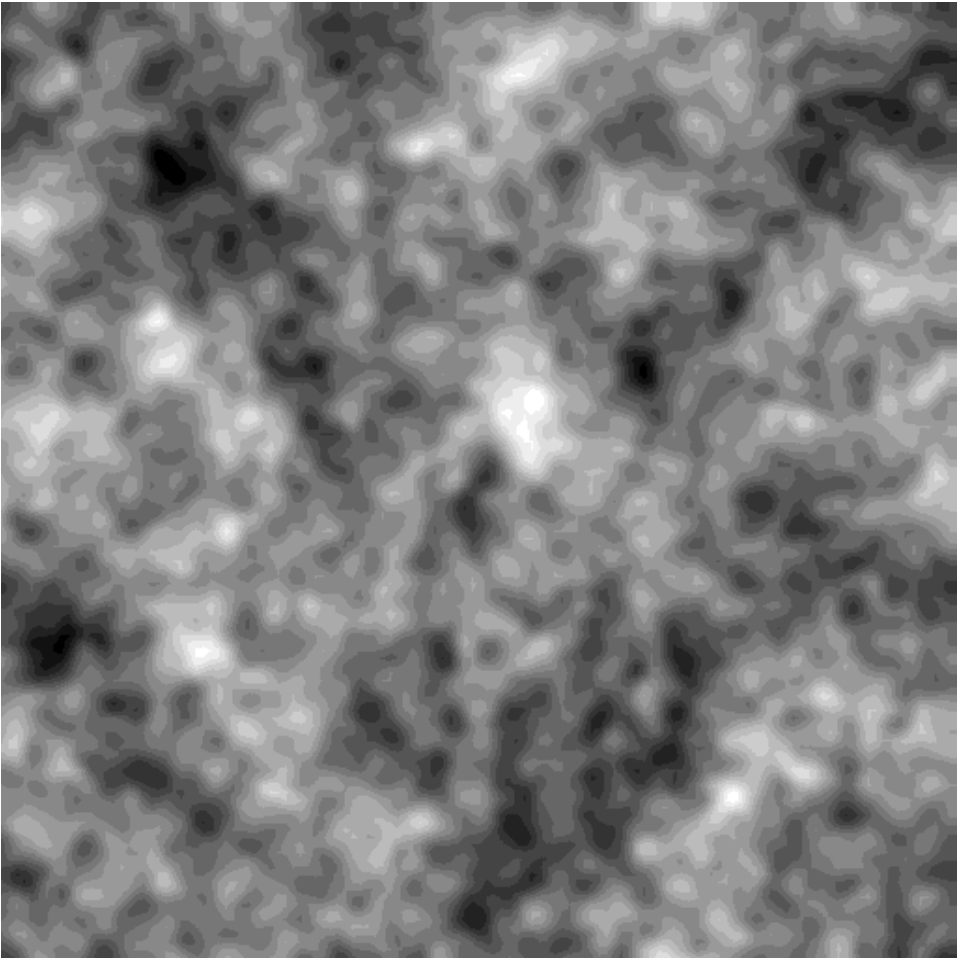

# Optimizing Observing Patterns in Cosmic Microwave Background Radiation Anisotropy Experiments

R.K. Schaefer and L. Piccirillo
Bartol Research Institute
University of Delaware
Newark, DE 19716

## Abstract

We consider several factors concerning the design of observing strategies in Cosmic Microwave Background (CMB) anisotropy experiments. First we consider the number of independent points on the sky one should observe given a fixed observing time. Given an assumed level of sky temperature fluctuations, $\sigma_{sky}$ and the presence of instrumental noise $s$ in units of K s$^{-1/2}$ and a total observing time $t_{obs}$, we find there is an optimum number of points in the sky to maximize the confidence level of a detection. This number is a function of the signal to noise ratio $R^2 = \sigma_{sky}^2 t_{obs}/s^2$. We verify this analytical result with a monte carlo simulation, which also includes the correlation between different positions in the sky. Furthermore, using an $n = 1$ spectrum of Gaussian fluctuations, we show that arranging the observing patterns along a (great) circle, a raster scan, or a randomized pattern yields results which are indistinguishable in Monte carlos.

# 1 Introduction

We have entered an era in which the long sought goal of detecting primordial temperature fluctuations has finally realized, beginning with the announcement of measured temperature fluctuations by the COBE team (Smoot, *et al.*, 1992). Since that announcement many other small scale experiments from the Earth (ground based and balloon borne) have also reported detections of anisotropy at levels similar to that found by COBE. Experiments from Earth also must deal with additional problems (*e.g.* atmospheric noise in ground based experiments and limited observing times (especially during balloon flights).

However, the difficulty of doing these experiments is well known. It makes sense therefore to study the elements necessary for designing the best experiment possible, and try to incorporate these into new experiments. We have begun a systematic study of the elements in experimental design and here we report on our results concerning observing strategies. In particular, the next section was motivated by our work with the TopHat balloon experiment (Cheng 1994a). We discuss the number of independent points on the sky which should be observed in order to get the most efficient detection. This is particularly relevant for balloon borne experiments, which have very limited observing time. In section 3 we consider the influence of choosing a particular pattern for the spots observed.

# 2 Choosing the Number of Spots

Typically one has many sources of noise in a cosmic microwave anisotropy experiment: atmospheric noise (emission and moving thermal gradients), thermal noise in the detector, and emission from other foreground sources (the earth, sun, moon, radio antennas, etc.) The ones with a non-random character (earth, sun, radio waves) can be minimized with a proper physical design of the instrument and pointing. The noise sources with a random character (atmosphere and detector noise) can be overcome by observing each spot for a long period of time and averaging. The experimenter can then spend a long time on each point until the random noises become insignificant. However, the observer usually has only a finite amount of time in which to observe. Thus we would like to know, is there an *optimal* number of spots on the sky to observe, given a fixed amount of observing time?

## 2.1 Level of Confidence for Detection: Uncorrelated Signals

Theories of galaxy formation predict relic temperature fluctuations which have some non-zero correlation over the sky. In theories not based on topological defects, it is natural to assume that the fluctuations are Gaussian. Indeed there are some observations of large scale structure which lend credence to this view (see, e.g. Strauss, et al. 1994; Moore, *et al.*, 1992). If we assume that the temperature fluctuations on the sky are Gaussian, but uncorrelated, we can analytically derive an optimum number of spots to observe, which depends only on the signal to noise ratio which we will call $R$. Of course, a more realistic model of temperature anisotropies includes the effect of temperature correlations, which we will address briefly in the section 2.2.

### 2.1.1 Maximum Likelihood Optimization

In an experiment one observes $N$ different spots on the sky resulting in a set of $N$ temperature fluctuation observations $\{\Delta T_i\}$. For every observation there is a given experimental noise with a distribution which can be characterized by its width $\sigma_i$. From this data set one wants to extract the intrinsic temperature variation in the CMB, called $\sigma_{sky}$. If we can assume both of these errors are Gaussian, (not true for all systematic noise), then we can analyse the data using the likelihood function $L(\sigma_{sky})$ (see e.g., Readhead, et al, 1987).

$$L(\sigma_{sky}) = \prod_{i=1}^{N} \left[2\pi(\sigma_i^2 + \sigma_{sky}^2)\right]^{-1/2} \exp\left[\frac{\Delta T_i^2}{2(\sigma_i^2 + \sigma_{sky}^2)}\right] \qquad (1)$$

The most likely value of $\sigma_{sky}$ is the one which maximizes the likelihood function. Once one has determined the maximal likelihood estimate of $\sigma_{sky}$, it is important to know how significant the detection is. There exists a "uniformly most powerful" test of the hypothesis that we have a detection. To be more precise, we take as our hypothesis, which we call $H$, that $\sigma_{sky}$, the maximal likelihood value, is an good measure of actual intrinsic fluctuations on the sky. As an alternative hypothesis, which we call $K$, we take $\sigma_{sky} = 0$.

We test the truth of these hypotheses with the likelihood ratio test. The likelihood ratio $\lambda$ is given by

$$\lambda = \frac{L(\sigma_{sky})}{L(0)} = \prod_{i=1}^{N} \left[ \frac{(\sigma_i^2)}{\sigma_i^2 + \sigma_{sky}^2} \right]^{1/2} \exp\left[ +\frac{\Delta T_i^2}{2\sigma_i^2} - \frac{\Delta T_i^2}{2(\sigma_i^2 + \sigma_{sky}^2)} \right] \qquad (2)$$

Our hypothesis ($H$) is acceptable if the ratio $\lambda$ is larger than some critical value $\lambda_c$. If this condition is met, we can specify the confidence level $1-\alpha$ of our detection, which is simply $1 - \alpha = P(\lambda > \lambda_c)|_H$ assuming our hypothesis $H$ is true. $\alpha$ is then the probability of falsely rejecting $H$, and is called a Type I error. However, $\lambda > 0$ for both hypotheses, so we must be careful that our test is powerful enough to reject $H$ when $K$ is true. This is called the power of the test $\beta$ and can be found from $\beta = P(\lambda < \lambda_c)|_K$, assuming $K$ is true. The possibility of accepting $H$ even though $K$ is true is called a Type II error and has probability $1 - \beta$. We note that the likelihood ratio test is a "uniformly most powerful" test, which means that for a given value of $\alpha$ the power of the test is maximized.

Our goal is to maximize the confidence level with respect to the number of observing points. However, the fact that the likelihood ratio maximizes the power of the test does not mean that the power cannot be extremely small for some tests. In fact maximizing only the confidence level by itself, gives a test with $\beta = 0$, putting us in the position that we could be very confident of the result of a meaningless test. The obvious way to proceed then is to maximize the confidence level *keeping the power of the test fixed*. This procedure was previously suggested by Vittorio and Mucciacia (1991) and also by Cottingham (D.C. Cottingham, private communication).

In order to be able to do the calculation analytically we make some simplifications. We take all of the measurement errors to be from the same Gaussian distribution with variance $\sigma_i^2 = NS^2$, where $S = s/\sqrt{t_{obs}}$ is the value of the noise integrated over the total observing time $t_{obs}$ of the experiment. Typically, $s$ is of the order of $\sim$ a few mK s$^{\frac{1}{2}}$. Note that this implies that we spend $t_{obs}/N$ amount of time observing each spot. In reality of course, there is some additional observing time lost to moving the antenna, but this is usually minimal and can be corrected for by subtracting the moving time from the total observing time.

To calculate the values $\alpha$ and $\beta$, it is simpler to calculate the distribution of $\lambda^*$ (which is the logarithm of the likelihood ratio, up to some factors)

$$\lambda^* = \sum_{i=1}^{N} \frac{x_i^2}{(NS^2 + \sigma_{sky}^2)NS^2} \qquad (3)$$

where $x_i$ is a Gaussian variable with variance $NS^2 + \sigma_{sky}^2$ under the hypothesis $H$ and variance $NS^2$ under $K$. The confidence level $1 - \alpha$ can then be found analytically from

$$P(\lambda > \lambda_c)|_H = \int\int \cdots \int \frac{dx_1 dx_2 \cdots dx_N}{[\sqrt{2\pi(NS^2 + \sigma_{sky}^2)}]^N} \exp\left[\frac{\sum_{i=1}^{N} x_i^2}{2(NS^2 + \sigma_{sky}^2)}\right] \qquad (4)$$

where the integral is over the volume where $\lambda > \lambda_c$. We can convert this integral to radial coordinates with the radial variable $W = \lambda NS^2/2$:

$$1 - \alpha = P(\lambda > \lambda_c)|_H = \frac{1}{\Gamma(N/2)} \int_{W_c}^{\infty} dW W^{N/2-1} e^{-W} \qquad (5)$$

where $W_c = \lambda_c NS^2/2$. We find the integral for the power of the test $\beta$:

$$\beta = P(\lambda < \lambda_c)|_K = \frac{1}{\Gamma(N/2)} \int_0^{W_c(1+R^2/N)} dW W^{N/2-1} e^{-W} \qquad (6)$$

where $R = \sigma_{sky}/S$ is the "signal to noise ratio" defined over the duration of the experiment.

We then choose a value for the power of the test, which fixes our value of $\lambda_c$ (or equivalently $W_c$) for any given number of points $N$. We then maximize the confidence level with respect to the number of points $N$. In figure 1 we display the confidence level as a function of the number of points for a fixed signal to noise ratio. We can see that the curve shows a broad maximum, which decreases sharply for very small $N$ and decreases slowly as we increase $N$. The maximum becomes broader and shifts to larger $N$ when we decrease the power of the test, becoming infinitely broad for $\beta = 0$. Thus if we do not fix the power of the test, we find the confidence level is maximized at $N = \infty$, with $\beta = 0$.

The values of $W_c$ are calculated numerically. The values of $W_c$ for reasonable values of the confidence level and power of the test will lie in the range $1/(1 + R^2/N)N/2 < W_c < N/2$, where the upper (lower) limit is the average value of $W$ under the hypothesis H (K).

The optimum number of points for a fixed power of the test then depends only on the signal to noise ratio $R$. In figure 2 we show how the optimum (maximum confidence level) number of points varies with the square of signal to noise ratio for several values of the power of the test. For $N > 10$ and $\beta = 0.95$, we can roughly say that $R^2 \approx 3N_{optimal}$. We can think of this result another way: if we want on optimal design with 95% power of the test, we should plan to integrate at each point until the signal to noise ratio is $\sim 3^{-1/2}$. This of course ignores the weak dependence on $N$, which of course is not valid at low $N$. For the COBE level of fluctuations (Bennett *et al.* 1994) $\sigma_{sky} = 30\mu K$ and a noise of $s = 1$ mK s$^{1/2}$, then each spot should be observed for about $10^4$ seconds.

We note that the optimum number in figure 2 scales roughly as $\sigma_{sky}^2$. Thus to really define an optimum experiment, one would like to have the answer first before designing the measurement. Perhaps the most straightforward use of figure 2 is for repeated scans of the same instrument. If one already has a detection, and would like to repeat the experiment to confirm that result, the obvious choice for $\sigma_{sky}$ is the already determined value. For first run experiments, we can estimate a value based on other experiments. Another approach one could use to estimate $\sigma_{sky}$ would be as follows. If one wants to test a particular theory one could use the prediction of that theory for $\sigma_{sky}$, If the theory is correct the experiment will result in the best detection.

In figure 2, we also plot the positions of different experiments based on the published detections. However, we can make estimates based on previous results. We note that this analysis ignores the correlation information, which can be used to get additional information from the data. Use of the correlation information will effectively increase the signal to noise ratio and will increase the value of the optimal number of points. We do not expect this to be a large correction to our formula for the optimal number of points (as we will argue in the next section). We are currently studying the effect of the correlation information and will report about it in a future publication. Generally, we find that based on our work, the current balloon experiments are reasonably close to optimal, but that many of the ground based experiments do not spend enough time observing each point. This is because there is considerably more atmospheric noise, even at relatively quiet sites like the south pole, at ground based sites compared to that seen from a balloon. This suggests that it makes sense to consider a fixed ground based observatory which can spend a good fraction of a day observing each point. Of course, the lim-iting factor with balloons is that there is only a finite amount of observing time, so that only about 10-20 points can be observed on a given flight. This can be somewhat overcome with a long duration balloon flight. The estimate of the experimental noise in the Boomerang experiment are for 3 $\mu$K noise (deBernardis, private communication) in each of 900 20 arcminute pixels in the sky. Assuming a COBE level of fluctuation, this would give Boomerang a signal to noise ratio of $R^2 = 9 \times 10^4$, which would be well off our plot. Based on estimates of the noise expected during the proposed long duration balloon TopHat experiment we place a line on the plot corresponding to the expected signal to noise ratio. We see that the experiment ought to observe between 100 and 200 points for an optimal detection.

Another way to think about this result is in terms of the ratio of cosmic variance to noise. In this simple Gaussian model the cosmic variance can be easily calculated. Rms sky fluctuations of level $\sigma_{sky}$ have a cosmic variance of $2^{1/4}\sigma_{sky}$ for each point. For an optimal detection, we want to have the ratio $\sigma_{sky}\sqrt{t_0/N}/s \approx 3^{1/2}$. The ratio of the cosmic variance per point to the noise per point is then $2^{1/4}\sigma_{sky}\sqrt{t_0/N}/s \approx 3^{1/2}2^{1/4} \approx 2$. This means an optimal detection has a noise level which is half of the cosmic variance level. We can take this as another way to define how long one should spend integrating each point. We need to get the noise down to the level of half of the cosmic variance, but it makes little sense to improve the noise limits below that level.

## 2.2 Level of Confidence for Detection: Correlated Signal

In realistic theories of galaxy formation the predicted temperature fluctuations are correlated on the sky. There is ample evidence of these correlations on large angular scales from COBE and FIRS (Smoot et al., 1992, Ganga *et al.*, 1994). One can ask if correlations would significantly alter the conclusions of the previous section. In order to answer this question, we have written a monte carlo code to simulate a typical differencing experiment on a simulated patch of sky containing scale invariant gaussian temperature fluctuations. The details of how we made the sky simulations can be found in the appendix.

On each sky map we perform the following experiment. We pick a set of points along a circle which fits in our square patch of simulated sky. At each point we perform a single beam differencing experiment (assuming a Gaussian shaped beam), by taking the difference between the temperature at two spots separated by a distance $\theta_{throw}$. If the underlying temperature fluctuations are Gaussian, the results of the differencing experiment will also be Gaussian, but with a larger ($\times 2$) variance. To each temperature difference $\Delta t_{i,CBR}$, we add a random Gaussian noise with a width of $\sigma = S^2 N$. Then we calculate the log of the likelihood ratio directly using equation (2). We then make a frequency of occurrence plot of the values of the log of the likelihood ratio. We have done this in figure 3 for 100,000 simulated experiments, each with 20 measurements along an observation circle.

To compare this correlated fluctuation map to the case of purely Gaussian fluctuations we plot the probability of getting a particular value of the likelihood ratio using equation (5) as a solid line in figure 3. From this plot we can see that the effect of correlations is very small as the monte carlo simulations match the uncorrelated Gaussian fluctuations quite closely. There is some skewing of the curve because the correlations tend to increase the variance over a pure Gaussian in some of the experiments, however the Gaussian approximation seems quite good for purposes of designing an experiment.

## 3   Spot Pattern

Since we have simulated sky patches we can also investigate the effect of different observing patterns on the sky. One can ask the question, should one choose a regular pattern on the sky, or should one choose points at random? We test this by comparing the results of simulated experiments with regular patterns (a circle and a raster scan) and a randomized choice of spots.

We perform the experiments on the simulated skies as described in the previous section. The observing pattern size is chosen to nearly fill the map. The diameter of the circle is picked to be nearly as big as the width of the map. We then observe at a fixed number of points on the circle (54). The raster scan is likewise chosen so that for the same number of points the raster pattern nearly fills the map. Lastly the random points are chosen with a uniform probability over the whole map. In figure 4 we show a comparison of the observing patterns on our maps.

We then calculate the most likely value of the temperature using the maximum of the likelihood function. We then find the average and the variance of these values over all of the simulations. The results are as follows. For 54 points in each of 10000 scans, we find that the temperature fluctuations have an rms average value of $12.64 \pm 2.00 \times 10^{-6}$ for the circle, $11.79 \pm 1.95 \times 10^{-6}$ for the raster scan and $11.78 \pm 1.97 \times 10^{-6}$ for the random scan. Since we put a level of fluctuations in the map which would give the a two beam chop level of $12.4 \times 10^{-6}$, all three patterns are consistent with this level within errors. The variance of each method is remarkably similar, which tells us that the three patterns are equivalent. We note that for each observation with the random pattern, a different random realization is used.

With small numbers ($\sim 1000$) of simulations we noticed that the random pattern occasionally would give much larger variances. We attribute this to the effect of the correlations in the temperature maps. When we checked the locations of the random pattern observations, we sometimes found them to be clumped in one particular area of the map. Thus the random patterns sometimes sample one particular region of a map more intensively than the regular patterns, so the longer wavelength correlations were not well sampled. This was the major question involved with using a random pattern. Averaging over a large number of simulations, ($\sim 10000$), we find that the random pattern does just as well as the regular pattern. We are interested in the random pattern as this type of strategy might be exploited to randomize (and thereby reduce) some systematic effects, (like atmospheric drifts, for example). This issue is something we are currently investigating.

Another concern, of course ,is that when designing the observing pattern on the sky, one must also avoid known sources of noise (the moon, the planets, the galaxy, *etc.*, so neither a completely regular nor a completely randomized pattern are probably realizable. In the design, we should aim to sample the long wavelength modes to get an accurate result.

Finally we point out we have only studied the effect of changing patterns which sample a particular piece of the sky uniformly. Of course one could imagine other scan patterns which vary the spatial frequencies being sampled.

**Acknowledgement**. We would like to acknowledge useful discussions with the Tophat experiment collaboration, especially helpful comments from Lyman Page. We would also like to thank the NSF and NASA for support.

## 4 Conclusions

We have explored some issues related to the design of observing patterns in cosmic microwave background anisotropy experiments. The first is the choice of the number of spots given a fixed amount of observing time. We find that in with Gaussian temperature fluctuations, the choice of the number of spots is dictated by the signal to noise ratio of the experiment $R$, where $R^2 = \sigma_{sky}^2 t_{obs}/s^2$, and $\sigma_{sky}$ is the intrinsic rms sky temperature fluctuations, $t_{obs}$ is the total available observing time in seconds, and $s$ is the experimental noise (in units of K $s^{1/2}$). The optimal number of points for a given $R$ and fixed power of the test can be found in figure 2. For a 95% power of the test, the number of spots to observe is $N \sim R^2/3$. By monte carlo simulations we show this result is not significantly affected by the presence of temperature correlations in the sky.

We have also shown that a regular pattern of observing positions is better than a random pattern in that the variance of results from different experiments is minimized with a regular pattern. The important factor seems to be obtaining a fair sample of longer wavelength correlations in the sky.

# 6 Figure Captions

Figure 1. The confidence level as a function of the number of points observed for a fixed power of the test (68%, 95%, and 99.7%). The maxima are fairly broad, and the sharpness of the maximum in confidence level increases as we increase the power of the test.

Figure 2. The number of points which maximizes the confidence level of the detection for a fixed power of the test (68%, 95%, and 99.7%). The x-axis is the signal to noise ratio of the experiment (see text) and the y-axis is the optimal number of independent observing positions, i. e., the value which maximizes the confidence level of the detection. The points are the configurations used in some experiments. The signal to noise ratio is calculated using the measured noise and quoted CMBR detection. For the proposed experiment TopHat, we used the expected sensitivity based on extrapolation from other balloon experiments. The acronyms of the experiments are MSAM (Cheng, *et al.*, 1994b), Sask, (Wollack, *et al.*, 1993), ARGO, (deBernardis, *et al.*, 1994), MAX (Devlin, *et al.*, 1994), IAB, (Piccirillo and Calisse, 1993), ACME, (Shuster *et al.*, 1993), SP89 (Meinhold and Lubin, 1989), Python, (Dragovan, *et al.*, 1994), Tenerife, (Hancock, *et al.*, 1994).

Figure 3. The probability of the likelihood function for an experiment with 20 points and $R = 5$, as a function of the (log of the) likelihood ratio. The points represent the results of $10^5$ mote carlo simulations on skies with a scale invariant spectrum. The curve is for an uncorrelated Gaussian sky. The effect of the correlations is to increase the likelihood of getting large signals in some of the maps, thus skewing the distribution.

Figure 4. The different observing patterns tested in our simulations. On the left we have the raster scan, in the center the circular pattern, and on the right, one example of a random pattern. The coordinates refer to map coordinates.

Figure 5. A grayscale pattern of a single patch of simulated sky with a scale invariant spectrum. The length of the square patch is 60 degrees and has been smoothed with a Gaussian of FWHM of 1 degree. The scale goes from black ($\delta T/T = -7.5 \times 10^{-5}$) to white ($\delta T/T = +7.5 \times 10^{-5}$).

## A  Monte Carlo simulations of experiments

We briefly describe the methods we use for setting up the simulated skies in our experiments. We simulate square patches of the sky using FFTs. For our purposes it is sufficient to use a gaussian random scale invariant spectrum on all scales. This is very similar to what is actually predicted in realistic models. In most the the currently popular models, the temperature fluctuations deviate slightly from a pure scale invariant spectrum on degree scales, which typify many Earth based experiments. However, these deviations are highly dependent on theoretical parameters, so a pure scale invariant spectrum is perhaps a good place to start. The power spectrum of the temperature fluctuations is often given in terms of the three dimensional Fourier space components. The temperature correlation function $C(\theta)$ between two directions on the sky specified by the unit vectors $\hat{\mathbf{e}}_1$ and $\hat{\mathbf{e}}_2$ can be written as

$$C(\theta) = \left\langle \frac{\delta T}{T}(\hat{\mathbf{e}}_1) \frac{\delta T}{T}(\hat{\mathbf{e}}_2) \right\rangle = \int d^3 k P_{rad}(k) \exp[i r_H \mathbf{k} \cdot (\hat{\mathbf{e}}_1 - \hat{\mathbf{e}}_2)], \quad (7)$$

where $\cos(\theta) = \hat{\mathbf{e}}_1 \cdot \hat{\mathbf{e}}_2$, and $P_{rad}(k)$ is the power spectrum of radiation fluctuations. For a scale invariant spectrum using this notation, $P_{rad}(k) \propto k^{-3}$. If the angle subtended by the map is not too large, one can approximate the patch of sky as being flat, and the vector $\hat{\mathbf{e}}_1 - \hat{\mathbf{e}}_2$ as perpendicular to the direction of observation, which we can take to be aligned with the $z$ axis. One can then integrate out the $k_z$ component (e.g. Doroshkevich, 1978; Vittorio and Mucciacia, 1991). The remaining integral is a two dimensional Fourier integral appropriate for our two dimensional simulation. We take the components $k_x$ and $k_y$ to be the conjugate variables to the map coordinates $\theta_x$ and $\theta_y$. The angular correlation function and the 2-dimensional radiation power spectrum of Vittorio and Mucciacia (1991) are then Mellin transform pairs. An alternative way of arriving at the same formulas is to use the more detailed method outlined in Bond and Efstathiou, (1987). They start by mapping the sphere into a disk whose radial coordinate $\omega$ is approximately equal to the polar angle of the sphere for small angles. They then define a transform pair $S(Q)$ and $C(\omega)$ (so for small angles $C(\Omega) \approx C(\theta)$). Their function $S(Q)$ is the same as the two dimensional radiation power spectrum $P_{rad}(q)$ in Mucciacia and Vittorio (1991), where $q = \sqrt{k_x^2 + k_y^2}$.

Using the 2 dimensional formalism of Vittorio and Mucciacia (1991), the temperature correlation function for a beam differencing experiment, which samples two spots with a Gaussian beam of width $\sigma$ an angle of $\theta$ apart, is given by

$$C(\theta, \sigma) = \int_0^\infty dq\, q P_{rad}(q) J_0(qr\theta) \exp[-q^2 r^2 \sigma^2] \qquad (8)$$

where $J_0$ is the Bessel function of order 0, $r = c/H_0$ is the horizon distance, and $P_{rad} \propto q^{-2}$ for a scale invariant spectrum in the Vittorio and Mucciacia definition. We can then calculate the expected average result of a single differencing experiment as

$$\left(\frac{\Delta T}{T}\right)^2 = 2C(0, \sigma) - 2C(\theta, \sigma) \qquad (9)$$

This formula is used to check that the monte carlos are giving the correct value of fluctuations.

For purposes of mapping we take a finite piece of the sky, so the Fourier transform becomes a Fourier series.

$$\frac{\delta T}{T}(\theta_x, \theta_y) = \sum_{n_x=1}^{N} \sum_{n_y=1}^{N} D(\mathbf{n}) exp(in_x \theta_x) exp(in_y \theta_y) \qquad (10)$$

The Fourier coefficients for a scale invariant spectrum are simply $D^2(\mathbf{n}) \propto \mathbf{n}^{-2}$, which are picked with a Guassian random distribution, and given a random phase. We note that in order to produce a map of temperature fluctuations which are real (not complex) certain reality conditions must be satisfied by the coefficients. (see, e.g. *Numerical Recipes*).

When simulating the experiments, the patch of sky must be smoothed with the beam pattern of the instrument. If we approximate this as a Gaussian, the convolution theorem tells us the whole map can be convolved with a Gaussian beam pattern by simply multiplying the Fourier coefficients by Gaussians before taking the Fourier transform.

Differences in the sky seen by the instrument are just differences of the temperatures on the maps. To simulate the experiments, we just choose an observing pattern and take the temperature differences of various locations on the map. The instrument noise is then simulated by adding Guassian noise to each "measurement" after it is taken.

For definiteness, our map is defined with a linear dimension of 60°. The experiment we simulate has a Gaussian beamwidth with a FWHM of 1° and a beamthrow of 1.6°. Using a purely scale invariant spectrum means that we can scale all the dimensions of the of the simulations together, with the warning that as we go to larger angular scales than this the distortions of the spherical map become worse. With our simulations, 60° already introduces distortions in the correlation function at the 10% level when compared to a true spherical correlation function. We feel that we are at the limiting size of a flat map and it does not make sense to use a flat map for a larger area than we consider here. For the simulations we use a 128 by 128 map of Fourier coefficients, so that we can do a large number in a short period of time. We show a sample map in figure 5.

We also would like to mention that the FFT technique does not reproduce the true correlation function expected in a patch of the sky in a scale invariant model (see, e.g., Bond and Efstathiou, 1987). This is because we have neglected the contribution of the wavelengths larger than the size of our map.